\documentclass[preprint2]{aastex}

\usepackage{epsfig}



\shorttitle{Weak gravitational studies of clusters}
\shortauthors{Dahle}
\slugcomment{ }

\begin{document}

\title{A compilation of weak gravitational lensing studies of clusters of galaxies.}

\author{H{\aa}kon Dahle \\
{\tt hdahle@astro.uio.no}, {\tt http://folk.uio.no/hdahle}}
\affil{Institute of Theoretical Astrophysics, University of Oslo, \\
P.O. Box 1029, Blindern, N-0315 Oslo, Norway}

\begin{abstract}

We present a list of clusters that have had their dark matter content measured using 
weak gravitational lensing. The list consists of 139 clusters, with weak lensing measurements 
reported in 64 different publications. Details are provided about the selection criteria 
and some basic properties of the sample, such as the redshift distribution.
An electronic, sortable version of this list with links to 
public database information on the clusters and publications is provided at \\
{\tt http://folk.uio.no/hdahle/WLclusters.html } 

\end{abstract}


\section{Introduction}

Since the first reported detection of weak gravitational shear 
produced by a massive cluster of galaxies (Tyson, Valdes, \& Wenk 1990), 
well over a hundred clusters have had their dark matter distribution mapped 
using this technique. Here, we present a compilation of published studies 
that have reported such measurements, either in the form of a map of the 
projected mass distribution in a cluster, or some quantity related to the cluster
mass, or both. 

\section{Selection criteria}

The studies are listed by cluster name in Table~\ref{tab:WLclusters}. 
Papers that combine strong and weak lensing data are also    
included in this table. The table does not include studies that derive 
the average mass of large ensembles of objects by measuring the mean gravitational shear
produced by these (e.g., Sheldon et al.\ 2001; Parker et al.\ 2005), 
as they do not provide results for individual objects. 

Also not included are $\sim 200$ ``shear-selected'' candidate clusters 
(e.g., Miyazaki et al.\ 2003; Hetterscheidt et al.\ 2005; Gavazzi \& Soucail 2006; 
Wittman et al.\ 2006; Schirmer et al.\ 2006), 
which have been detected directly from their weak lensing effect. 
Such cluster samples will contain some fraction of spurious detections, 
arising both from projections of multiple lesser structures along the line of sight, and 
from random alignment of background galaxies, resembling a lensing signal 
from a real cluster (e.g., White, van Waerbeke, \& Mackey 2002; 
Hennawi \& Spergel 2005). Hence, as would be expected, the currently reported cluster 
candidates range from peaks in the reconstructed projected density distribution with no obvious optical 
and X-ray counterparts, to well-established overdensities in 3D space with measured 
spectroscopic redshifts and corresponding extended X-ray emission. 

The compilation of Table~\ref{tab:WLclusters} does not include weak gravitational lensing 
studies based on measurements of magnification bias (Broadhurst, Taylor, \& Peacock 1995), 
since this is a rather different technique than shear-based mass measurements, and 
has only been applied to a small number of clusters so far. 

A few studies of superclusters have been published (Kaiser et al.\ 1998; 
Gray et al.\ 2002; Gavazzi et al.\ 2004; Dietrich et al.\ 2005; Jee et al.\ 2006). 
For these systems, a separate entry is given in Table~\ref{tab:WLclusters} 
for each of their constituent clusters. 

It should be noted that most of the clusters listed in Table~\ref{tab:WLclusters} have several 
alternative designations, and the naming covention adopted may differ between various publications, and
also between these and the corresponding entry in databases such as the NASA/IPAC 
Extragalactic Database (NED). Generally, the naming convention most 
commonly used in the literature has been adopted, and NED links are provided in the 
online version of Table~\ref{tab:WLclusters} to refer the reader to alternative designations. 

\section{Properties of the sample}

\begin{figure*}
\centering\epsfig{file=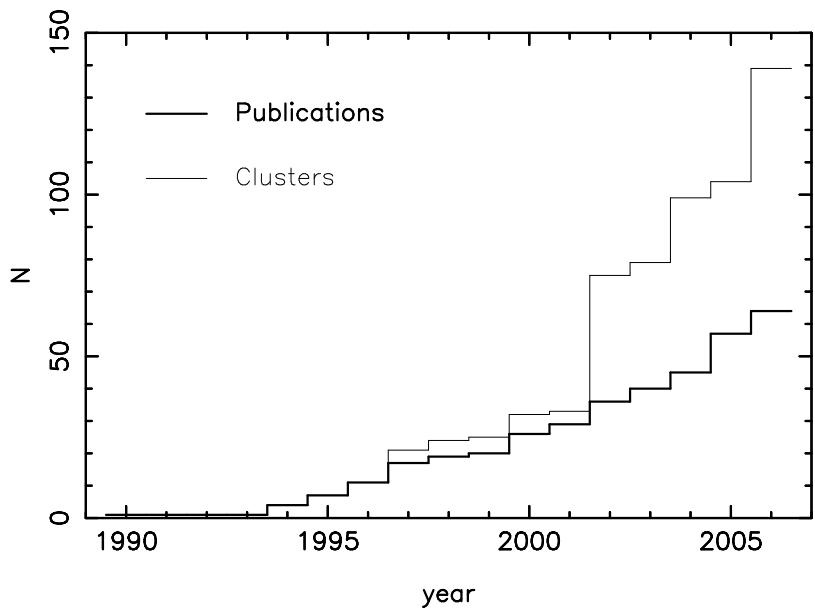,width=8cm,angle=-90}
\caption[Redshift distribution.]
{The cumulative number of clusters with weak lensing data, and the cumulative count of papers listed in Table~\ref{tab:WLclusters}, as a function of the year of publication.}
\label{fig:cumul}
\end{figure*}

The list provided in Table~\ref{tab:WLclusters} contains a total of 139 clusters, with 
weak lensing data reported in 64 separate publications. 
Figure~\ref{fig:cumul} illustrates how these have accumulated over time 
(with a clear tendency for more clusters per publication in recent years).  
The majority of these clusters 
were originally identified in optical cluster surveys (e.g., Abell, Corwin, \& Olowin 1989, 
Gonzalez et al.\ 2001), while most of the remaining clusters were found by optical 
followup of X-ray surveys such as the Einstein Medium-Sensitivity Survey (Gioia et al.\ 1990) 
or the ROSAT All-Sky Survey (Tr{\"u}mper 1993). 

\begin{figure*}
\centering\epsfig{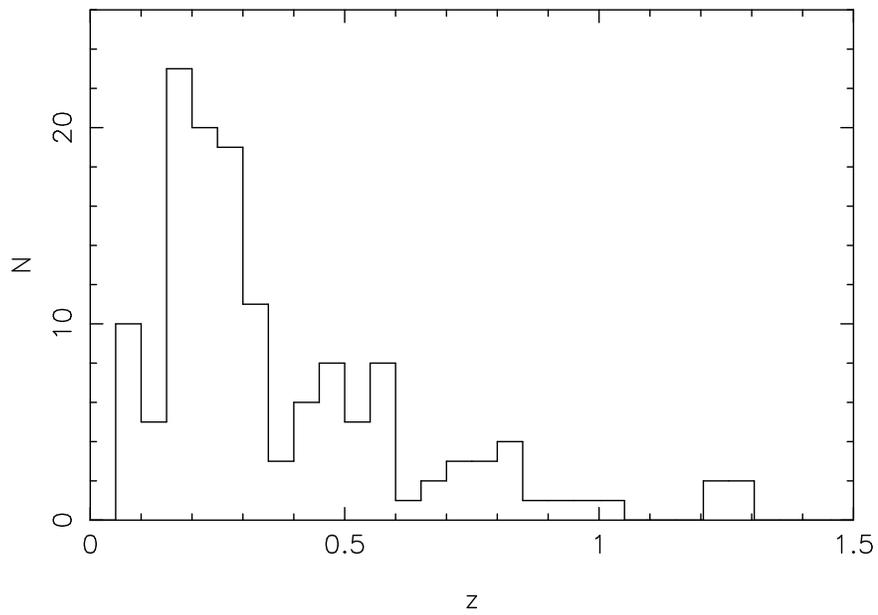}
\caption[Redshift distribution.]
{The redshift distribution of galaxy clusters with weak lensing measurements, in bins of width $\Delta z = 0.05$.} 
\label{fig:zdistr}
\end{figure*}

The redshift distribution of the 
published weak lensing cluster sample is shown in Figure~\ref{fig:zdistr}. 
None of the clusters are at $z<0.05$, where the lensing efficiency is very low, and 
only five clusters are at $z=1$ or higher, where weak lensing studies are only feasible
using the {\it Hubble Space Telescope} (HST). About 75\% of the clusters are at $z<0.5$. The 
distribution in Figure~\ref{fig:zdistr} has a strong peak around $z \sim 0.2-0.3$, where clusters will act most efficiently 
as lenses, given the redshift distribution of background galaxies in typical ground-based 
imaging data.   
  
The current version of this compilation does not tabulate any cluster mass values provided 
in the respective publications listed in Table~\ref{tab:WLclusters}. Considerable caution 
is warranted when combining such data from different publications: Firstly, the angular 
diameter distances assumed when translating the measured shear into cluster mass depends 
on the assumed cosmological model, and most studies older than $\sim 5$ years adopted cosmological 
parameters that differ significantly from the current ``concordance cosmology''. 
Secondly, our knowledge of the redshift distribution of faint galaxies has improved significantly 
since the early papers (particularly with the advent of photometric redshift measurements of 
faint galaxies in deep HST images), and this will again affect the mass values through the source distance 
estimates. Thirdly, the methodology for measuring gravitational shear has evolved considerably, 
although the most popular method (``KSB+''; see Kaiser, Squires, \& Broadhurst 1995; 
Luppino \& Kaiser 1997; Hoekstra et al.\ 1998) was developed already in the mid-1990s, and 
has been shown through simulations to produce results sufficiently accurate for measurements 
of weak lensing by clusters (e.g., Heymans et al.\ 2006). 

Finally, it should be noted that the reported mass values are often not directly comparable
to each other, as some are 2D estimates of the projected mass in a cylinder, e.g., using the ``aperture 
densitometry'' estimator (Kaiser et al.\ 1994), while others report an estimated velocity 
dispersion, or an estimated cluster 
mass within a 3D volume. The latter quantities are typically derived by fitting the observed tangential 
component of the shear as a 
function of cluster radius to a spherically symmetric theoretical model 
such as a singular isothermal sphere or an NFW model (Navarro, Frenk, \& White 1997). 
The relation between 2D and 3D masses can be calculated for any of these 
mass models, but the reliability of the results will be sensitive to sub-clustering and other 
departures from such simple mass models. 
   
\acknowledgments

HD acknowledges support from the Research Council of Norway, including a postdoctoral research fellowship. This research has made use of the NASA/IPAC Extragalactic Database (NED) which
is operated by the Jet Propulsion Laboratory, California Institute of 
Technology, under contract with the National Aeronautics and Space Administration.

\begin{deluxetable}{lcl}
\tabletypesize{\small}
\tablewidth{10cm}
\tablecaption{Weak gravitational lensing measurements of galaxy clusters
  \label{tab:WLclusters}}
\tablehead{
\colhead{Cluster} &
\colhead{Redshift} & 
\colhead{ } 
\\
\colhead{name} & \colhead{$z$ \tablenotemark{a} } & \colhead{Publication(s)} \\ 
}
\startdata
\objectname{1E 0657-558}                   & 0.296 & Clowe et al.\             2004 \\
\nodata                                    &\nodata& Bradac et al.\            2006 \\
\nodata                                    &\nodata& Clowe et al.\             2006b\\
\objectname{3C 254}                        & 0.736 & Wold et al.\              2002 \\
\objectname{3C 295}                        & 0.460 & Smail et al.\             1997 \\
\nodata                                    &\nodata& Wold et al.\              2002 \\
\objectname{3C 324}                        & 1.206 & Smail and Dickinson       1995 \\
\objectname{3C 334}                        & 0.555 & Wold et al.\              2002 \\
\objectname{3C 336}                        & 0.927 & Fort et al.\              1996 \\
\nodata                                    &\nodata& Bower and Smail           1997 \\
\objectname{Abell 0022}                    & 0.142 & Cypriano et al.\          2004 \\
\objectname{Abell 0068}                    & 0.255 & Dahle et al.\             2002 \\
\nodata                                    &\nodata& Smith et al.\             2005 \\
\nodata                                    &\nodata& Dahle                     2006 \\
\objectname{Abell 0085}                    & 0.055 & Cypriano et al.\          2004 \\
\objectname{Abell 0115}                    & 0.197 & Dahle et al.\             2002 \\
\nodata                                    &\nodata& Dahle                     2006 \\
\objectname{Abell 0141}                    & 0.230 & Dahle et al.\             2002 \\
\objectname{Abell 0209}                    & 0.206 & Dahle et al.\             2002 \\
\nodata                                    &\nodata& Smith et al.\             2005 \\
\objectname{Abell 0222}                    & 0.213 & Dietrich et al.\          2005 \\
\objectname{Abell 0223}                    & 0.207 & Dietrich et al.\          2005 \\
\objectname{Abell 0267}                    & 0.231 & Dahle et al.\             2002 \\
\nodata                                    &\nodata& Smith et al.\             2005 \\
\nodata                                    &\nodata& Dahle                     2006 \\
\objectname{Abell 0383}                    & 0.187 & Smith et al.\             2005 \\
\objectname{Abell 0520}                    & 0.199 & Dahle et al.\             2002 \\
\nodata                                    &\nodata& Dahle                     2006 \\
\objectname{Abell 0586}                    & 0.171 & Dahle et al.\             2002 \\
\nodata                                    &\nodata& Cypriano et al.\          2005 \\
\nodata                                    &\nodata& Dahle                     2006 \\
\objectname{Abell 0611}                    & 0.288 & Dahle                     2006 \\
\objectname{Abell 0665}                    & 0.182 & Dahle et al.\             2002 \\
\nodata                                    &\nodata& Dahle                     2006 \\
\objectname{Abell 0697}                    & 0.282 & Dahle et al.\             2002 \\
\nodata                                    &\nodata& Dahle                     2006 \\
\objectname{Abell 0773}                    & 0.217 & Dahle et al.\             2002 \\
\nodata                                    &\nodata& Smith et al.\             2005 \\
\nodata                                    &\nodata& Dahle                     2006 \\
\objectname{Abell 0781}                    & 0.298 & Dahle                     2006 \\
\objectname{Abell 0851}                    & 0.407 & Seitz et al.\             1996 \\
\nodata                                    &\nodata& Smail et al.\             1997 \\
\objectname{Abell 0901}                    & 0.170 & Gray et al.\              2002 \\
\objectname{Abell 0902}                    & 0.160 & Gray et al.\              2002 \\
\objectname{Abell 0914}                    & 0.193 & Dahle et al.\             2002 \\
\objectname{Abell 0922}                    & 0.190 & Dahle et al.\             2002 \\
\objectname{Abell 0959}                    & 0.286 & Dahle et al.\             2002 \\
\objectname{Abell 0963}                    & 0.206 & Dahle et al.\             2002 \\
\nodata                                    &\nodata& Smith et al.\             2005 \\
\nodata                                    &\nodata& Dahle                     2006 \\
\objectname{Abell 1351}                    & 0.328 & Dahle et al.\             2002 \\
\objectname{Abell 1437}                    & 0.135 & Cypriano et al.\          2004 \\
\objectname{Abell 1451}                    & 0.171 & Cypriano et al.\          2004 \\
\objectname{Abell 1553}                    & 0.165 & Cypriano et al.\          2004 \\
\objectname{Abell 1576}                    & 0.299 & Dahle et al.\             2002 \\
\nodata                                    &\nodata& Dahle                     2006 \\
\objectname{Abell 1650}                    & 0.084 & Cypriano et al.\          2004 \\
\objectname{Abell 1651}                    & 0.085 & Cypriano et al.\          2004 \\
\objectname{Abell 1664}                    & 0.128 & Cypriano et al.\          2004 \\
\objectname{Abell 1682}                    & 0.234 & Dahle et al.\             2002 \\
\nodata                                    &\nodata& Dahle                     2006 \\
\objectname{Abell 1689}                    & 0.183 & Tyson, Valdes and Wenk    1990 \\
\nodata                                    &\nodata& Tyson and Fischer         1995 \\
\nodata                                    &\nodata& Clowe and Schneider       2001 \\
\nodata                                    &\nodata& King et al.\              2002 \\
\nodata                                    &\nodata& Bardeau et al.\           2005 \\
\nodata                                    &\nodata& Broadhurst et al.\        2005 \\
\nodata                                    &\nodata& Limousin et al.\          2006 \\
\objectname{Abell 1705}                    & 0.297 & Dahle et al.\             2002 \\
\objectname{Abell 1722}                    & 0.326 & Dahle et al.\             2002 \\
\objectname{Abell 1758}                    & 0.279 & Dahle et al.\             2002 \\
\nodata                                    &\nodata& Dahle                     2006 \\
\objectname{Abell 1763}                    & 0.223 & Dahle et al.\             2002 \\
\nodata                                    &\nodata& Smith et al.\             2005 \\
\nodata                                    &\nodata& Dahle                     2006 \\
\objectname{Abell 1835}                    & 0.253 & Clowe and Schneider       2002 \\
\nodata                                    &\nodata& Dahle et al.\             2002 \\
\nodata                                    &\nodata& Smith et al.\             2005 \\
\nodata                                    &\nodata& Dahle                     2006 \\
\objectname{Abell 1914}                    & 0.171 & Dahle et al.\             2002 \\
\nodata                                    &\nodata& Dahle                     2006 \\
\objectname{Abell 1942}                    & 0.224 & Erben et al.\             2000 \\
\objectname{Abell 1995}                    & 0.321 & Dahle et al.\             2002 \\
\objectname{Abell 2029}                    & 0.077 & Menard, Erben and Mellier 2003 \\
\nodata                                    &\nodata& Cypriano et al.\          2004 \\
\objectname{Abell 2104}                    & 0.153 & Dahle et al.\             2002 \\
\nodata                                    &\nodata& Cypriano et al.\          2004 \\
\objectname{Abell 2111}                    & 0.229 & Dahle et al.\             2002 \\
\nodata                                    &\nodata& Dahle                     2006 \\
\objectname{Abell 2163}                    & 0.203 & Squires et al.\           1997 \\
\nodata                                    &\nodata& Cypriano et al.\          2004 \\
\objectname{Abell 2204}                    & 0.152 & Clowe and Schneider       2002 \\
\nodata                                    &\nodata& Dahle et al.\             2002 \\
\nodata                                    &\nodata& Cypriano et al.\          2004 \\
\nodata                                    &\nodata& Dahle                     2006 \\
\objectname{Abell 2218}                    & 0.176 & Squires et al.\           1996a\\
\nodata                                    &\nodata& Smail et al.\             1997 \\
\nodata                                    &\nodata& Smith et al.\             2005 \\
\objectname{Abell 2219}                    & 0.226 & Dahle et al.\             2002 \\
\nodata                                    &\nodata& Hoekstra et al.\          2002 \\
\nodata                                    &\nodata& Smith et al.\             2005 \\
\nodata                                    &\nodata& Dahle                     2006 \\
\objectname{Abell 2261}                    & 0.224 & Dahle et al.\             2002 \\
\nodata                                    &\nodata& Dahle                     2006 \\
\objectname{Abell 2345}                    & 0.177 & Dahle et al.\             2002 \\
\nodata                                    &\nodata& Cypriano et al.\          2004 \\
\objectname{Abell 2384}                    & 0.094 & Cypriano et al.\          2004 \\
\objectname{Abell 2390}                    & 0.228 & Squires et al.\           1996b\\
\nodata                                    &\nodata& Dahle                     2006 \\
\nodata                                    &\nodata& Hicks et al.\             2006 \\
\objectname{Abell 2426}                    & 0.098 & Cypriano et al.\          2004 \\
\objectname{Abell 2537}                    & 0.295 & Dahle et al.\             2002 \\
\objectname{Abell 2552}                    & 0.300 & Dahle                     2006 \\
\objectname{Abell 2597}                    & 0.085 & Cypriano et al.\          2004 \\
\objectname{Abell 2631}                    & 0.273 & Dahle                     2006 \\
\objectname{Abell 2744}                    & 0.308 & Smail et al.\             1997 \\
\objectname{Abell 2811}                    & 0.108 & Cypriano et al.\          2004 \\
\objectname{Abell 2843}                    & 0.560 & Smail et al.\             1997 \\
\objectname{Abell 3667}                    & 0.056 & Joffre et al.\            2000 \\
\objectname{Abell 3695}                    & 0.089 & Cypriano et al.\          2004 \\
\objectname{Abell 3739}                    & 0.165 & Cypriano et al.\          2004 \\
\objectname{Abell 3856}                    & 0.138 & Cypriano et al.\          2004 \\
\objectname{Abell 3888}                    & 0.153 & Cypriano et al.\          2004 \\
\objectname{Abell 3984}                    & 0.181 & Cypriano et al.\          2004 \\
\objectname{Abell 4010}                    & 0.095 & Cypriano et al.\          2004 \\
\objectname{Abell S910}                    & 0.311 & Smail et al.\             1997 \\
\objectname{Cl 0016+1609}                  & 0.541 & Smail et al.\             1995 \\
\nodata                                    &\nodata& Smail et al.\             1997 \\
\nodata                                    &\nodata& Clowe et al.\             2000 \\
\nodata                                    &\nodata& Hicks et al.\             2006 \\
\objectname{Cl 0024+1654}                  & 0.390 & Bonnet, Mellier and Fort  1994 \\
\nodata                                    &\nodata& Smail et al.\             1997 \\
\nodata                                    &\nodata& Kneib et al.\             2004 \\
\objectname{Cl 0303+1706}                  & 0.420 & Kaiser et al.\            1998 \\
\nodata                                    &\nodata& Gavazzi et al.\           2004 \\
\nodata                                    &\nodata& Smail et al.\             1997 \\
\objectname{Cl 0413-6559}                  & 0.510 & Smail et al.\             1997 \\
\objectname{Cl 0957+561}                   & 0.355 & Dahle, Maddox and Lilje   1994 \\
\nodata                                    &\nodata& Fischer et al.\           1997 \\
\objectname{Cl 1601+4253}                  & 0.539 & Smail et al.\             1997 \\
\objectname{Cl 1604+4304}                  & 0.897 & Margoniner et al.\        2005 \\
\nodata                                    &\nodata& Umetsu and Futamase       2000 \\
\objectname{E 1821+643}                    & 0.297 & Wold et al.\              2002 \\
\objectname{J105511.6-050416}              & 0.680 & Wittman et al.\           2003 \\
\objectname{J1312.5+7252}                  & 0.550 & Dahle et al.\             2003 \\
\objectname{J234624.00+004358.8}           & 0.333 & Wittman et al.\           2001 \\
\objectname{LCDCS 057}                     & 0.473 & Clowe et al.\             2006a\\
\objectname{LCDCS 110}                     & 0.579 & Clowe et al.\             2006a\\
\objectname{LCDCS 130}                     & 0.704 & Clowe et al.\             2006a\\
\objectname{LCDCS 172}                     & 0.697 & Clowe et al.\             2006a\\
\objectname{LCDCS 173}                     & 0.749 & Clowe et al.\             2006a\\
\objectname{LCDCS 188}                     & 0.456 & Clowe et al.\             2006a\\
\objectname{LCDCS 198}                     & 0.960 & Clowe et al.\             2006a\\
\objectname{LCDCS 252}                     & 0.550 & Clowe et al.\             2006a\\
\objectname{LCDCS 275}                     & 0.807 & Clowe et al.\             2006a\\
\objectname{LCDCS 340}                     & 0.480 & Clowe et al.\             2006a\\
\objectname{LCDCS 430}                     & 0.424 & Clowe et al.\             2006a\\
\objectname{LCDCS 504}                     & 0.794 & Clowe et al.\             2006a\\
\objectname{LCDCS 531}                     & 0.636 & Clowe et al.\             2006a\\
\objectname{LCDCS 541}                     & 0.541 & Clowe et al.\             2006a\\
\objectname{LCDCS 567}                     & 0.465 & Clowe et al.\             2006a\\
\objectname{LCDCS 634}                     & 0.483 & Clowe et al.\             2006a\\
\objectname{LCDCS 849}                     & 0.588 & Clowe et al.\             2006a\\
\objectname{LCDCS 853}                     & 0.763 & Clowe et al.\             2006a\\
\objectname{LCDCS 925}                     & 0.520 & Clowe et al.\             2006a\\
\objectname{LCDCS 952}                     & 0.496 & Clowe et al.\             2006a\\
\objectname{MG 2016+112}                   & 1.004 & Benitez et al.\           1999 \\
\nodata                                    &\nodata& Clowe, Trentham and Tonry 2001 \\
\objectname{MS 0302.5+1717}                & 0.425 & Kaiser et al.\            1998 \\
\nodata                                    &\nodata& Gavazzi et al.\           2004 \\
\objectname{MS 0302.7+1658}                & 0.424 & Kaiser et al.\            1998 \\
\nodata                                    &\nodata& Gavazzi et al.\           2004 \\
\objectname{MS 0440.5+0204}                & 0.190 & Smail et al.\             1997 \\
\objectname{MS 0451.6-0305}                & 0.550 & Clowe et al.\             2000 \\
\objectname{MS 0906.5+1110}                & 0.180 & Hicks et al.\             2006 \\
\objectname{MS 1008.1-1224}                & 0.301 & Lombardi et al.\          2000 \\
\nodata                                    &\nodata& Athreya et al.\           2002 \\
\objectname{MS 1054.4-0321}                & 0.823 & Luppino and Kaiser        1997 \\
\nodata                                    &\nodata& Clowe et al.\             2000 \\
\nodata                                    &\nodata& Hoekstra et al.\          2000 \\
\nodata                                    &\nodata& Jee et al.\               2005a\\
\objectname{MS 1137.5+6624}                & 0.782 & Clowe et al.\             2000 \\
\objectname{MS 1224.7+2007}                & 0.327 & Fahlman et al.\           1994 \\
\objectname{MS 1358.4+6245}                & 0.328 & Hicks et al.\             2006 \\
\nodata                                    &\nodata& Hoekstra et al.\          1998 \\
\objectname{MS 1512.4+3647}                & 0.372 & Hicks et al.\             2006 \\
\objectname{MS 1621.5+2640}                & 0.426 & Hicks et al.\             2006 \\
\objectname{MS 2053.7-0449}                & 0.583 & Clowe et al.\             2000 \\
\nodata                                    &\nodata& Hoekstra et al.\          2002 \\
\objectname{MS 2137.3-2353}                & 0.313 & Gavazzi et al.\           2003 \\
\nodata                                    &\nodata& Gavazzi                   2005 \\
\objectname{RDCS 1252.9-2927}              & 1.237 & Lombardi et al.\          2005 \\
\objectname{RX J0152.7-1357}               & 0.831 & Huo et al.\               2004 \\
\nodata                                    &\nodata& Jee et al.\               2005b\\
\objectname{RX J0437.1+0043}               & 0.285 & Dahle                     2006 \\
\objectname{RX J0439.0+0715}               & 0.230 & Dahle                     2006 \\
\objectname{RX J0848.6+4453}               & 1.270 & Jee et al.\               2006 \\
\objectname{RX J0848.9+4452}               & 1.261 & Jee et al.\               2006 \\
\objectname{RX J1157.3+3336}               & 0.214 & Dahle et al.\             2002 \\
\nodata                                    &\nodata& Dahle                     2006 \\
\objectname{RX J1347.5-1145}               & 0.451 & Fischer and Tyson         1997 \\
\nodata                                    &\nodata& Bradac et al.\            2005 \\
\nodata                                    &\nodata& Kling et al.\             2005 \\
\objectname{RX J1532.9+3021}               & 0.345 & Dahle et al.\             2002 \\
\objectname{RX J1716.4+6708}               & 0.813 & Clowe et al.\             2000 \\
\objectname{RX J1720.1+2637}               & 0.164 & Dahle et al.\             2002 \\
\nodata                                    &\nodata& Dahle                     2006 \\
\objectname{RX J2129.6+0005}               & 0.235 & Dahle et al.\             2002 \\
\nodata                                    &\nodata& Dahle                     2006 \\
\objectname{Zwicky 2089}                   & 0.230 & Dahle et al.\             2002 \\
\nodata                                    &\nodata& Dahle                     2006 \\
\objectname{Zwicky 3146}                   & 0.291 & Dahle                     2006 \\
\objectname{Zwicky 5247}                   & 0.229 & Dahle et al.\             2002 \\
\nodata                                    &\nodata& Dahle                     2006 \\
\objectname{Zwicky 5768}                   & 0.266 & Dahle                     2006 \\
\objectname{Zwicky 7160}\tablenotemark{b}  & 0.258 & Smail et al.\             1995 \\
\nodata                                    &\nodata& Dahle et al.\             2002 \\
\nodata                                    &\nodata& Dahle                     2006 \\
\nodata                                    &\nodata& Hicks et al.\             2006 \\
\objectname{Zwicky 7215}                   & 0.290 & Dahle                     2006 \\
\enddata
\tablecomments{A sortable electronic version of this table with links to NED and ADS is available at \\
{\tt http://folk.uio.no/hdahle/WLclusters.html}.} 
\tablenotetext{a}{--- Most cluster redshift values are taken from the NASA/IPAC Extragalactic Database (NED), with the exception of redshifts for \objectname{Abell 914}, \objectname{Abell 1351}, \objectname{Abell 1576}, \objectname{Abell 1722}, and \objectname{Abell 1995}, which have tabulated redshift values from Irgens et al.\ (2002), and \objectname{Abell 2552}, which has a redshift value from Ebeling et al.\ (2000).}
\tablenotetext{b}{--- Alternative designation: \objectname{MS 1455.0+2232}.}
\end{deluxetable}

{
}

\end{document}